# Effect of Volume and Temperature on the Global and Segmental Dynamics in Polypropylene Glycol and 1,4-polyisoprene

C.M. Roland[1], R. Casalini[1,2], and M. Paluch[3]

[1]Naval Research Laboratory  
Chemistry Division, Code 6120  
Washington, D.C. 20375-5342

[2]George Mason University  
Chemistry Department  
Fairfax, VA 22030

[3]Silesian University  
Institute of Physics  
Uniwersytecka 4, 40-007  
Katowice, Poland

ABSTRACT:

Published dielectric relaxation measurements on polypropylene glycol and 1,4-polyisoprene are analyzed to determine the relative effect that thermal energy and volume have on the temperature dependence of the normal mode relaxation times, and compare this to their effect on the temperature dependence of the local segmental relaxation times. We find that for both polymers at temperatures well above $T_g$, both relaxation modes are governed more by thermal energy than by volume, although the latter's contribution is not negligible. Such a result is consistent with an assumption underlying models for polymer viscoelasticity, such as the Rouse and tube models, that the friction coefficient governing motions over large length scales can be identified with the local segmental friction coefficient. We also show that relaxation data for both the segmental and the normal mode superimpose, when expressed as a function of the product of the temperature and the volume, the latter raised to a power. This scaling form arises from an inverse power form for the intermolecular potential. The value of the exponent on the volume for these two polymers indicates a relatively "soft" potential.

**Introduction**

The rheology of polymeric materials has obvious practical significance, and has been the focus of much research over the past half-century. The slow dynamics of entangled molecules is usually interpreted in terms of reptation[1], which emphasizes the disparity in the spatial constraints on transverse motions, as opposed to motion along the chain backbone. The Doi-Edwards tube model [2] provides a theoretical framework for the reptation idea and, with various refinements over the past twenty-five years, has become predominant in the field[3,4]. The tube model describes the entangled dynamics as Rouse chains moving in a network of topological constraints. There are two species-dependent parameters, the local (Rouse) friction coefficient, $\varsigma$, and a parameter characterizing the entanglements. For both the tube and Rouse models, the temperature dependence of the chain dynamics time is contained in $\varsigma$. Since this is the same friction coefficient governing local segmental relaxation [5,6], the implication is that the local modes and the chain modes should have the same temperature dependence. (This expectation is implicit in master curves for the dynamic properties of polymers which extend from the rubbery plateau into the glassy zone.) However, as shown by the breakdown of time-temperature superpositioning in the softening zone[7-11], the local dynamics and the chain modes do not have the same temperature dependence. Within the confines of the tube and Rouse models, the discrepancy lies in the identification of $\varsigma$ with the friction coefficient for segmental relaxation.

Herein we analyze published dielectric spectroscopy data, to investigate local segmental relaxation (dielectric α-process) and the chain dynamics (dielectric normal mode process) in polypropylene glycol (PPG) and polyisoprene (PI). Both polymers are barely entangled: The PPG has a molecular weight equal to 4.0 kg/mol, with entanglements reputed to be abetted by transient coupling via hydrogen-bonding of the chain ends.[12]. For the PI, $M_w$ = 11.1 kg/mol, which is almost a factor of two larger than the entanglement molecular weight.[8] PPG and PI are both type-A polymers, having a permanent dipole moment parallel to the chain; consequently, global motions of the chains backbone are dielectrically active. Dielectric relaxation times for the normal mode, $\tau_N$, and for local segmental relaxation, $\tau_\alpha$, have been measured for PPG[13] and PI[14], as a function of both temperature and pressure. Although the focus of the tube and Rouse models is on the rheological response, mechanical and dielectric measurements probe the chain dynamics in an analogous fashion. Thus, while the shape of the respective relaxation spectra can differ[4], the temperature and pressure dependences of the two measurements are the same.[8,10,14-16]

By combining the dielectric results for these two materials with equation of state (*PVT*) data, the volume-dependence of the relaxation times can be obtained. More specifically, we can assess the relative degree to which thermal energy and volume govern the respective temperature dependences of $\tau_\alpha$ and $\tau_N$.

Previously this type of analysis has been carried out only for the segmental relaxation process.[17-19,20] For small-molecule, van der Waals liquids, the contribution to $\tau_\alpha(T)$ from temperature and volume is nearly the same. On the other hand, for polymers, generally temperature exerts a more dominant influence on the segmental relaxation. However, the effect of volume is not negligible. and becomes more important for more flexible chain polymers, such as the siloxanes.[21] For strongly hydrogen-bonded materials, temperature tends to become overwhelmingly the dominant control parameter.

In the present work, we extend the analysis of temperature and volume effects to consider the normal mode. We also employ a recently proposed scaling of the relaxation times, based on an inverse power form for the intermolecular repulsive potential.[22] We find that for PPG and PI well above $T_g$, $\tau_\alpha$ and $\tau_N$ are governed similarly by *T* and *V*. Accordingly, master curves for each are obtained, using the same scaling exponent for the two relaxation times. Our findings give strong support to a basic tenet of the tube and Rouse models, namely the use of the local segmental friction factor in the description of the global (chain) dynamics.

**Results**

**Polyproylene glycol.** We have previously reported the segmental and normal mode relaxation times for PPG, measured at pressures from ambient to as high as 1.2 GPa, at five temperatures from 258 to 313K.[13] The relaxation times were defined as the inverse of the angular frequency associated with the maximum in the dielectric loss, which are close to the average relaxation times, if the peaks are not broad. For this molecular weight of PPG (4 kg/mol), there is a substantial difference between $\tau_\alpha$ and $\tau_N$. However, since the segmental relaxation times are more sensitive to both temperature[23] and pressure[13,24] than are the normal mode relaxation times, conditions can be found for which the two relaxation processes can be measured simultaneously. We convert the isotherms (isothermal $\tau$ as a function of *P*) from ref. [13] to a function of volume by using the equation state for PPG[13]

$$V(T,P) = \left(0.9852 + 7.2 \times 10^{-4} T + 4.7 \times 10^{-7} T^2\right)\left(1 - 0.0894 \ln\left(1 + \frac{P}{171\exp(-0.0052T)}\right)\right) \quad (1)$$

where the specific volume, V, is in mL/g, and the units of pressure and temperature are MPa and Centigrade, respectively. The results are plotted in Fig. 1.[25] It can be seen that the relaxation times for either process are not uniquely defined by the volume; thermal energy obviously exerts an influence. However, $\tau_\alpha$ and $\tau_N$ are also changing with pressure for any fixed temperature, which means that volume also plays a role.

We can quantify the relative effects of temperature and volume from the ratio of the thermal expansivity measured at a constant value of the relaxation time, $\alpha_\tau = V^{-1} \left.\frac{\partial V}{\partial T}\right|_\tau$, to the isobaric value of the thermal expansion coefficient, $\alpha_P = V^{-1} \left.\frac{\partial V}{\partial T}\right|_P$.[18] This ratio, $|\alpha_\tau|/\alpha_P$ (since $\alpha_\tau < 0$, the absolute value is the quantity of interest), measures how much the volume would have to change (by a variation of pressure) in order for the relaxation time to remain constant when the temperature is varied. Thus, $|\alpha_\tau|/\alpha_P = 1$ if temperature (thermal energy) and volume exert an equivalent effect on the temperature dependence of the relaxation times, whereas the ratio will be much larger than one if temperature is the more dominant control variable.[18] The expansivity ratio is related to the apparent activation enthalpies (slope of the Arrhenius curves) at constant volume, $H_V$ and constant pressure, $E_P$, as $\alpha_\tau/\alpha_P = -H_V/(E_P - H_V)$[26]. The quantity $H_V/E_P$ is another measure of the relative effect of temperature and volume on $\tau(T)$.[17]

For PPG at ambient pressure near $T_g$ (~202K), $|\alpha_\tau|/\alpha_P = 2.0 \pm 0.4$.[13], which means that for low temperatures, thermal energy exerts roughly twice the influence that the specific volume does, in determining the change of $\tau_\alpha$ with temperature. This result is consistent with early work of Williams.[27] Near $T_g$, the chain modes are too slow to measure dielectrically. In order to compare temperature and volume effects on both $\tau_\alpha$ and $\tau_N$ at the same temperature, we calculate the expansivity ratios for the two processes at a higher temperature, 288 K, and elevated pressure, $P$ = 186 MPa. These particular conditions avoid the need to extrapolate any of the data in Fig. 1. From eq. 1, $\alpha_P = 4.72 \times 10^{-4}$ K$^{-1}$. The isochronal thermal expansion coefficient is calculated at 288K and pressures whereby $\tau_N$ is a fixed value = 0.01 s, yielding $\alpha_{\tau N}$ = -1.30 (± 0.1) ×10$^{-3}$ K$^{-1}$. At T = 288 K and P = 186 MPa, the segmental relaxation time = 2.0×10$^{-5}$ s. The volume expansivity at which $\tau_\alpha$ is fixed at this value is calculated to be $\alpha_{\tau,\alpha}$ = -1.25 (± 0.2) ×10$^{-3}$ K$^{-1}$. As summarized in Table 1, the ratio of these quantities is then obtained as $|\alpha_{\tau,N}|/\alpha_P$ = 2.8 ± 0.2 for the normal mode, and $|\alpha_{\tau,\alpha}|/\alpha_P$ = 2.6 ± 0.3 for the segmental mode.

The expansivity ratio for the segmental mode, $|\alpha_{\tau,\alpha}|/\alpha_P$ = 2.6, is larger than the value at $T_g$[13], indicating the relative contribution from thermal energy increases at higher temperatures. In fact, we find there is a consistent trend of slightly increasing $|\alpha_{\tau,\alpha}|/\alpha_P$ with decreasing $\tau_\alpha$. Of greater interest herein is the finding that within the experimental error, local segmental relaxation and the normal mode are governed equivalently by T and V: $|\alpha_{\tau,\alpha}|/\alpha_P \approx |\alpha_{\tau,N}|/\alpha_P \approx 2.7$. That is, roughly three-quarters of the increase in either relaxation time with increasing temperature is a direct result of greater thermal energy, the remaining increase in τ due to the accompanying volume expansion. It is interesting that even though both the temperature- and pressure-dependences of the two relaxation times differ, at least in the regime where $\tau_N \sim 0.01$ s and $\tau_\alpha$ is about 3 decades shorter, the relative effects of thermal energy and volume are the same for the two processes. The larger variation in $\tau_\alpha$ with change in T or V is due to greater sensitivity of the local segmental dynamics to these variables, not to a difference in their relative effect on $\tau_\alpha$. The relatively weak effect of V on $\tau_\alpha$(T) and $\tau_N$(T) is ostensibly at odds with free volume interpretations of polymer dynamics; however, the free volume and the specific volume are not equivalent. In fact, the former can even change at fixed V.[5]

**1,4-polyisoprene.** Floudas and Reisinger[14] reported isotherms for both $\tau_\alpha$ and $\tau_N$ of PI at two temperatures, for pressures up to 350 MPa. The reported Vogel-Fulcher fits to these data are plotted in Figure 2 as a function of the specific volume, the latter determined from the equation of state for PI[14]

$$V(T,P) = \left(1.0943 + 6.293 \times 10^{-4}T + 6.231 \times 10^{-7}T^2\right)\left(1 - 0.0894 \ln\left(1 + \frac{P}{202 \exp(-0.004653T)}\right)\right) \quad (2)$$

From Fig. 2 it is clear that temperature and volume both affect the temperature dependence of the relaxation times. In the manner done for PPG, we calculate the isobaric and isochronal thermal expansion coefficients. To avoid any extrapolation, we choose T = 263 K and P = 200 MPa, for which $\alpha_P = 3.589 \times 10^{-4}$ K$^{-1}$. The relaxation times at this temperature and pressure are $\tau_N = 0.01$ s and $\tau_\alpha = 3.1 \times 10^{-6}$ s, for which the corresponding isochronal expansivities are calculated to be $\alpha_{\tau,N} = -1.08 (\pm 0.1) \times 10^{-3}$ K$^{-1}$ and $\alpha_{\tau,\alpha} = -1.18 (\pm 0.15) \times 10^{-3}$ K$^{-1}$. From these data, we calculate the expansivity ratios listed in Table 1; within the experimental error, $|\alpha_{\tau,N}|/\alpha_P = |\alpha_{\tau,\alpha}|/\alpha_P \approx 3.2$. Thus, similar to the case of PPG, the relative effect of temperature and volume is the same for the two relaxation times. Also note that at this relatively high temperature, thermal energy has a stronger influence on the relaxation times than does volume. Floudas and Reisinger[14] reached a similar conclusion specifically for the α-relaxation at lower temperatures, near $T_g$.

**Scaling of the relaxation times.** There is no accurate theoretical prediction for the combined *T*- and *P*-dependences of either the segmental or the normal mode relaxation times. And, while the $\tau_\alpha(T)$ data are known to conform to the Vogel-Fulcher (or WLF) equation up to *ca.* 1.3 × $T_g$, the origin of this behavior is uncertain, with different models, based variously on free volume[28-30] or entropy[31-33], showing limited success. For the pressure-dependence of $\tau_\alpha$, the situation is worse, with different models yielding various forms for $\tau_\alpha(P)$. Concerning the chain dynamics, there is only the inference from the Rouse and tube models that $\tau_N$ should follow the temperature and pressure dependences of segmental relaxation.

We recently proposed a generalized scaling of $\tau_\alpha$ data obtained for both temperature and pressure. It is based on an inverse power-law, $\varphi(r) \propto r^{-3\gamma}$, for the repulsive potential, where r is the intermolecular separation and $\gamma$ is a material parameter.[34] The underlying idea is that the liquid structure is primarily determined by repulsive forces, with the attractive forces serving as a background potential which hold the molecules together.[35,36] With this form, all thermodynamic properties of the material can be expressed as a function of the variable $T^{-1}V^{-\gamma}$.[37] We have found that dynamic quantities related to the glass-transition relaxation, such as $\tau_\alpha$, the ionic conductivity, and the viscosity of monomeric glass-formers, can also be expressed as a single function of the variable $T^{-1}V^{-\gamma}$.[22] Moreover, the magnitude of $\gamma$ is correlated with the relative contribution of temperature and volume to the local dynamics. For the extreme cases, hard spheres (volume dominated) and thermally activated dynamics, $\gamma = \infty$ and 0 respectively. For various glass-formers, including van der Waals molecules, associated liquids, and polymers, we find $0.1 < \gamma < 9$, paralleling the magnitude of $|\alpha_\tau|/\alpha_P$.

If this scaling applies to the global dynamics, which has not heretofore been shown, our results suggest that (i) for each polymer, $\tau_\alpha$ and $\tau_N$ should exhibit the same scaling, since $|\alpha_{\tau,\alpha}|/\alpha_P \approx |\alpha_{\tau,N}|/\alpha_P$, and (ii) since the relative effect of temperature is roughly comparable for PI and PPG, their respective scaling exponents should be similar.

In Figure 3, $\tau_\alpha$ and $\tau_N$ for the two polymers are displayed as a function of $T^{-1}V^{-\gamma}$. For both modes, the same value of $\gamma$ causes the relaxation times for various $T$ and $P$ to fall on a single curve. Over as much as 8 decades, the superpositioning is quite good. We also find that the scaling exponents, $\gamma = 2.5 \pm 0.35$ for PPG and $= 3.0 \pm 0.15$ for PI, are close, in accord with the approximate equivalence of their respective $|\alpha_\tau|/\alpha_P$. These values of $\gamma$ are fairly small, reflecting the softness of the intermolecular potential for these polymers.

**Summary**

We find that for two barely-entangled polymers, at temperatures for which the normal mode relaxation time is in the range from 0.01 to 0.1 s, the respective temperature dependences of the segmental and normal modes are governed very similarly by thermal energy and volume. This result is congruent with an implicit assumption of the Rouse and tube models, that the relevant friction coefficient for the global dynamics can be identified with the friction coefficient for local segmental relaxation. However, our results ostensibly contradict the fact that the segmental and chain modes have different temperature[7-10] and pressure dependences[13,14]. However, the same relative influence from thermal energy and volume on $\tau_\alpha$ and $\tau_N$ does not require that the magnitude of the change induced in either relaxation time must be the same. For segmental relaxation, the effects may be amplified, for example in the manner described by the coupling model.[11]

Similar to literature results for segmental relaxation near $T_g$ in other polymers, we find for both PPG and PI that thermal energy exerts a more significant effect on the temperature-dependence of $\tau_\alpha$ than does volume. This implies a relatively "soft" intermolecular potential, and indeed, superpositioning of the relaxation times is achieved using a small value of the scaling exponent. This scaling exponent comes from an inverse power form for the intermolecular potential. Finally, we note that our analysis was carried out for temperatures well above $T_g$. At these high temperatures, the segmental relaxation times are ~ 10 μs. When $\tau_\alpha$ becomes so short, the temperature dependences of the segmental and chain modes are expected to be comparable[10]. However, as seen in the present data, differences in pressure dependences are still observed.

**Acknowledgements**

This work was supported by the Office of Naval Research. We thank G.M. Poliskie and K.L. Ngai for useful comments.

**References**

(1)  de Gennes, P. G. *Journal of Chemical Physics* **1971**, *55*, 572-579.

(2)  Doi, M.; Edwards, S. F. *The Theory of Polymer Dynamics*; Clarendon: Oxford, 1986.

(3)  Lodge, T. P.; Rotstein, N. A.; Prager, S. *Advances in Chemical Physics* **1990**, *79*, 1-132.

(4)  Watanabe, H. *Progress in Polymer Science* **1999**, *24*, 1253-1403.

(5)  Ferry, J. D. *Viscoelastic Properties of Polymers*; Wiley: New York, 1980.

(6)  Yamakawa, H. *Modern Theory of Polymer Solutions*; Harper and Row: New York, 1971.

(7)  Plazek, D. J. *Journal of Physical Chemistry* **1965**, *69*, 3480-&.

(8)  Santangelo, P. G.; Roland, C. M. *Macromolecules* **1998**, *31*, 3715-3719.

(9)  Plazek, D. J.; Chay, I. C.; Ngai, K. L.; Roland, C. M. *Macromolecules* **1995**, *28*, 6432-6436.

(10) Roland, C. M.; Ngai, K. L.; Santangelo, P. G.; Qiu, X. H.; Ediger, M. D.; Plazek, D. J. *Macromolecules* **2001**, *34*, 6159-6160.

(11) Ngai, K. L.; Plazek, D. J. *Rubber Chemistry and Technology* **1995**, *68*, 376-434.

(12) Fleischer, G.; Helmstedt, M.; Alig, I. *Polymer Communications* **1990**, *31*, 409-411.

(13) Roland, C. M.; Psurek, T.; Pawlus, S.; Paluch, M. *Journal of Polymer Science Part B-Polymer Physics* **2003**, *41*, 3047-3052.

(14)     Floudas, G.; Reisinger, T. *Journal of Chemical Physics* **1999**, *111*, 5201-5204.

(15)     McCrum, N. G.; Read, B. E.; Williams, G. *Anelastic and Dielectric Effects in Polymeric Solids*; Dover: New York, 1991.

(16)     Floudas, G.; Gravalides, C.; Reisinger, T.; Wegner, G. *Journal of Chemical Physics* **1999**, *111*, 9847-9852.

(17)     Williams, G. In *Dielectric Spectroscopy of Polymeric Materials*; Runt, J. P.; Fitzgerald, J. J., Eds.; American Chemical Society: Washington D.C., 1997.

(18)     Ferrer, M. L.; Lawrence, C.; Demirjian, B. G.; Kivelson, D.; Alba-Simionesco, C.; Tarjus, G. *Journal of Chemical Physics* **1998**, *109*, 8010-8015.

(19)     Paluch, M.; Casalini, R.; Roland, C. M. *Physical Review B* **2002**, *66*, 092202.

(20)     Roland, C.M.; Paluch, M.; Pakula, T.; Casalini, R. Phil. Mag. B 2004, in press

(21)     Paluch, M.; Casalini, R.; Patkowski, A.; Pakula, T.; Roland, C. M. *Physical Review E* **2003**, *68*.

(22)     Casalini, R.; Roland, C. M. *Physical Review Letters* **2003**, *submitted*.

(23)     Ngai, K. L.; Schonhals, A.; Schlosser, E. *Macromolecules* **1992**, *25*, 4915-4919.

(24)     Andersson, S. P.; Andersson, O. *Macromolecules* **1998**, *31*, 2999-3006.

(25)     The data for PPG at 283K in ref. 13 had significant scatter; thus, these were re-measured.

(26)     Casalini, R.; Roland, C. M. *Journal of Chemical Physics* **2003**, *119*, 4052-4059.

(27)     Williams, G. *Transactions of the Faraday Society* **1965**, *61*, 1564-&.

(28)     Cohen, M. H.; Grest, G. S. *Physical Review B* **1979**, *20*, 1077-1098.

(29)     Paluch, M.; Casalini, R.; Roland, C. M. *Physical Review E* **2003**, *67*, 021508.

(30)   Corezzi, S.; Capaccioli, S.; Casalini, R.; Fioretto, D.; Paluch, M.; Rolla, P. A. *Chemical Physics Letters* **2000**, *320*, 113-117.

(31)   Adam, G.; Gibbs, J. H. *Journal of Chemical Physics* **1965**, *43*, 139.

(32)   Casalini, R.; Paluch, M.; Fontanella, J. J.; Roland, C. M. *Journal of Chemical Physics* **2002**, *117*, 4901-4906.

(33)   Casalini, R.; Capaccioli, S.; Lucchesi, M.; Rolla, P. A.; Corezzi, S. *Physical Review E* **2001**, *6303*, 031207.

(34)   Hoover, W. G.; Ross, M. *Contemporary Physics* **1971**, *12*, 339-356.

(35)   Weeks, J. D.; Chandler, D.; Andersen, H. C. *Journal of Chemical Physics* **1971**, *54*, 5237-5247.

(36)   Shell, M. S.; Debenedetti, P. G.; La Nave, E.; Sciortino, F. *Journal of Chemical Physics* **2003**, *118*, 8821-8830.

(37)   March, N. H.; Tosi, M. P. *Introduction to Liquid State Physics*; World Scientific: Singapore, 2002.

Table 1. Results for polypropylene glycol and 1,4-polyisoprene

|     | T    | P       | $\tau_N$ | $\tau_\alpha$         | $-\alpha_{\tau,N}/\alpha_P$ | $-\alpha_{\tau,\alpha}/\alpha_P$ | $\gamma$       |
|-----|------|---------|----------|-----------------------|-----------------------------|----------------------------------|----------------|
| PPG | 258K | 186 MPa | 0.01 s   | $2.0\times10^{-5}$ s  | $2.8 \pm 0.2$               | $2.6 \pm 0.3$                    | $2.5 \pm 0.35$ |
| PI  | 283K | 200 MPa | 0.1 s    | $3.1\times10^{-6}$ s  | $3.0 \pm 0.3$               | $3.3 \pm 0.4$                    | $3.0 \pm 0.15$ |

**Figure Captions**

Figure 1. Segmental (hollow symbols) and normal mode (half-filled symbols) relaxation times for PPG as a function of the specific volume. The measurements[13,25] were carried out at the indicated temperatures at varying pressures.

Figure 2. Segmental (hollow symbols) and normal mode (half-filled symbols) relaxation times for PI as a function of the specific volume. The measurements[14] were carried out at the two indicated temperatures at pressures varying from 0.1 to 350 MPa for $\tau_N$ and 150 to 350 MPa for $\tau_\alpha$.

Figure 3. Relaxation times from Figs. 1 and 2 as a function of the quantity $T^{-1}V^{-\gamma}$, where $\gamma = 2.5$ and 3.0 for PPG and PI, respectively. For PPG, $T$ = 258 (○), 268 (◁), 283 (△), 293 (▽), 303 (◇), and 313 K (□), and for PI, $T$ = 283 (▽) and 298 K (○).

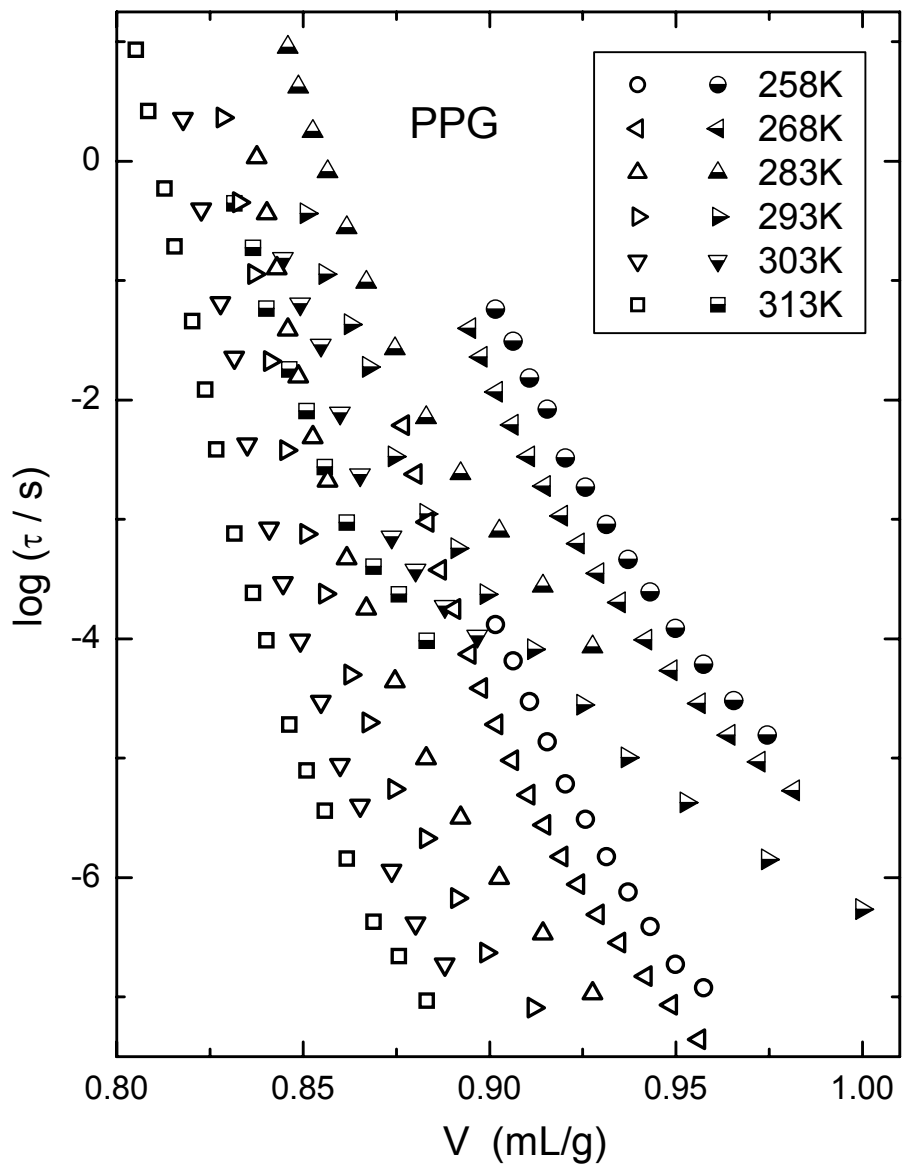

figure 1

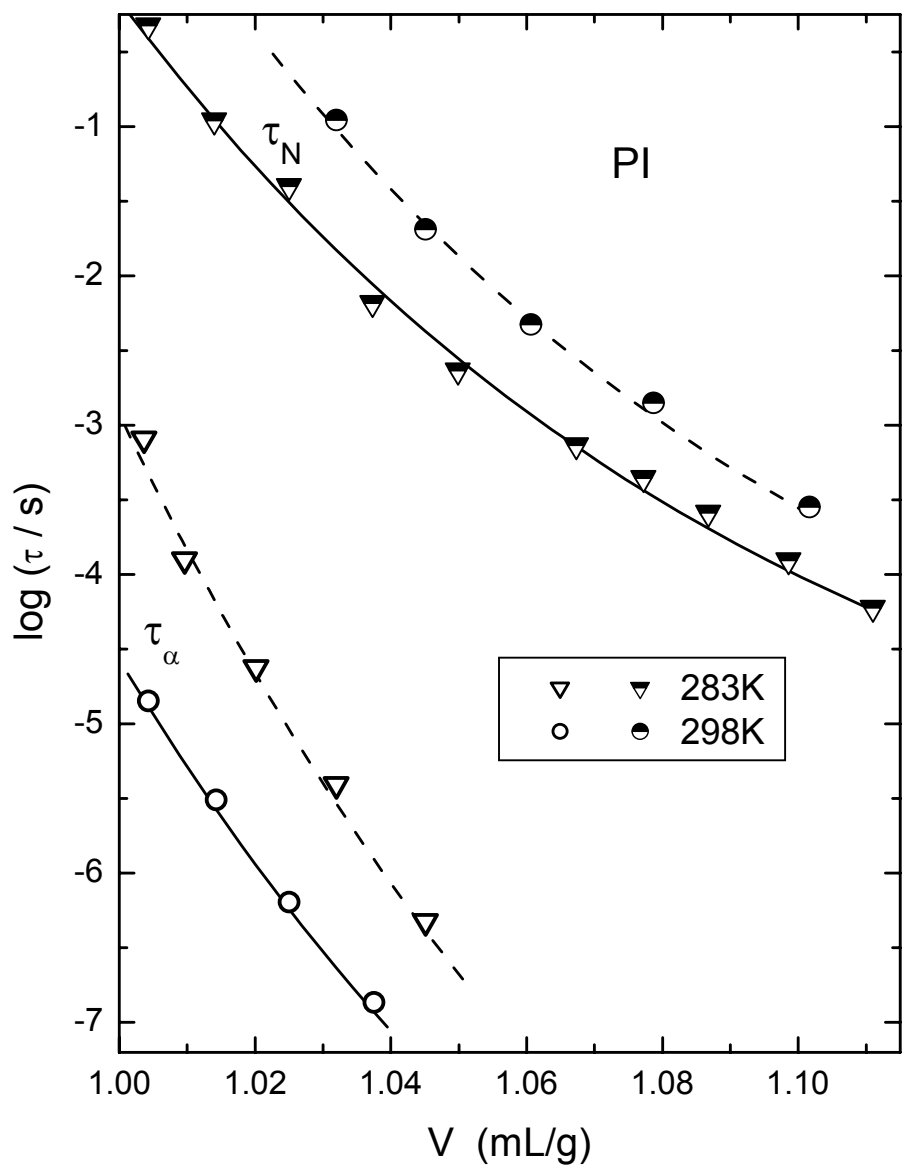

figure 2

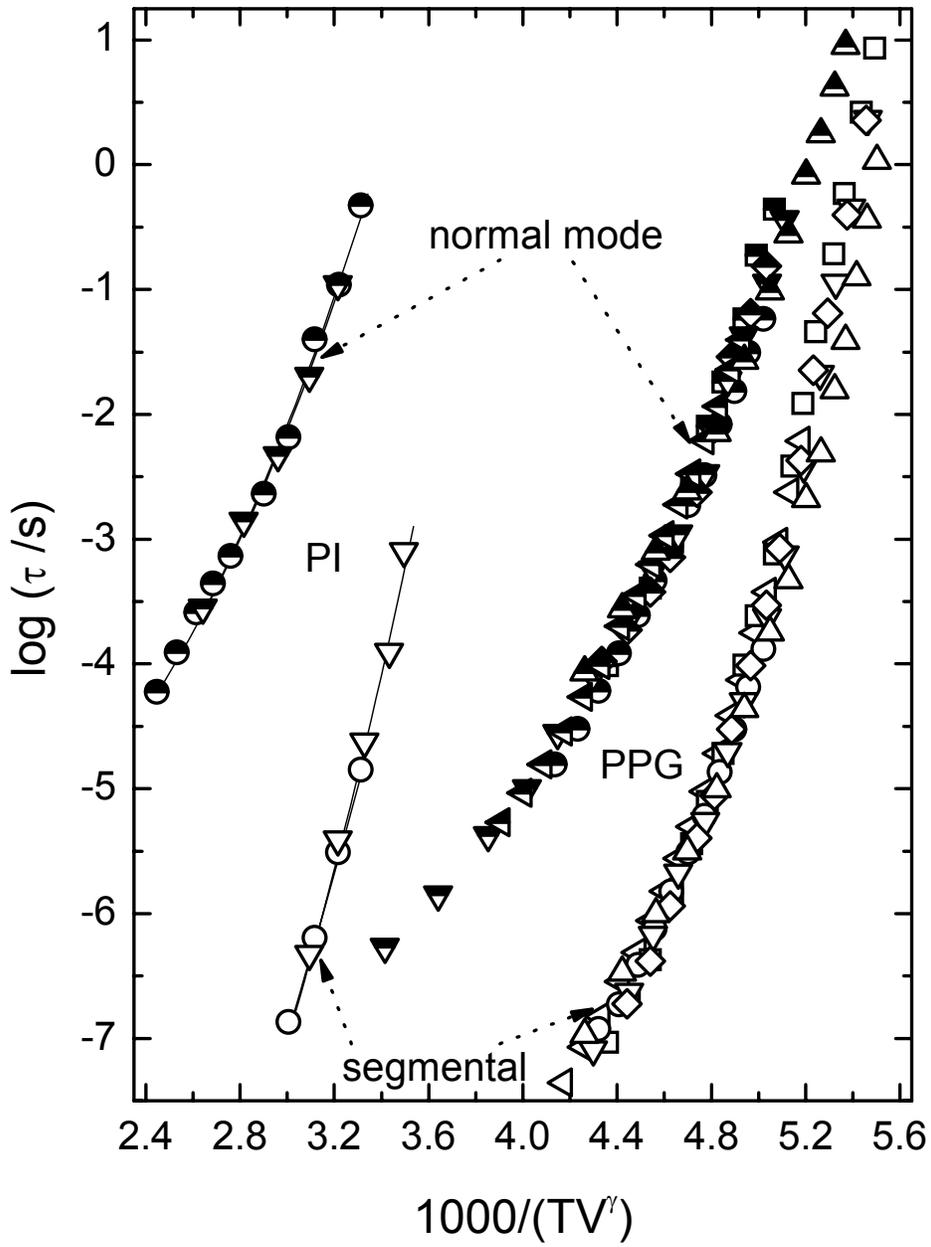

figure 3